\documentclass[twocolumn,showpacs,amssymb,aps,prl]{revtex4}
\usepackage[T1]{fontenc}
\usepackage{amsmath}
\usepackage{amssymb}
\usepackage{graphicx}
\newcommand{\be}{\begin{equation}}
\newcommand{\ee}{\end{equation}}

\begin{document}

\title{Toroidal trapped surfaces and isoperimetric inequalities}
\author{Janusz Karkowski}\email{janusz.karkowski@uj.edu.pl}
\affiliation{
Institute of Physics, Jagiellonian University, \L ojasiewicza 11,  30-438 Krak\'ow, Poland}

\author{Patryk Mach}\email{patryk.mach@uj.edu.pl}
\affiliation{
Institute of Physics, Jagiellonian University, \L ojasiewicza 11,  30-438 Krak\'ow, Poland}

\author{Edward Malec}\email{malec@th.if.uj.edu.pl}
\affiliation{
Institute of Physics, Jagiellonian University, \L ojasiewicza 11,  30-438 Krak\'ow, Poland}

\author{Niall \'{O} Murchadha}\email{niall@ucc.ie}
\affiliation{
Department of Physics, University College Cork, Cork, Ireland}

\author{Naqing Xie}\email{nqxie@fudan.edu.cn}
  \affiliation{School of Mathematical Sciences, Fudan University, Shanghai 200433, China}

\date{\today}

\begin{abstract}
We analytically construct an infinite number of trapped toroids in spherically symmetric Cauchy hypersurfaces of the Einstein equations. We focus on initial data which represent ``constant density stars'' momentarily at rest. There exists an infinite number of constant mean curvature tori, but we also deal with more general configurations. The marginally trapped toroids have been found analytically and numerically; they are   unstable. The topologically toroidal trapped surfaces appear in a finite region   surrounded by the Schwarzschild horizon.
\end{abstract}

\maketitle

\section{Introduction}

There exists numerical evidence that black hole horizons can have nonspherical topologies at an early stage of their development \cite{Masaru, HKWWST, STW}. It has been speculated that apparent horizons also possess a nontrivial topology (\cite{Masaru, Husa96} and references therein). Recently, quantitative criteria were found that guarantee the appearance of toroidal trapped surfaces \cite{KHX} on an initial Cauchy hypersurface. In this paper, we explicitly construct trapped surfaces of toroidal topology in spherically symmetric geometries generated by ``a constant density star'' \cite{NOM}. Remarkably, they can appear only in strongly curved geometries, when the spherical apparent horizon is already present. In this environment, one can find, under some additional conditions, a region filled with trapped toroids.

In Section 2 we define our coordinate tori inside a constant density star. Section 3 is dedicated to the derivation of relations between parameters --- the toroidal radii and the asymptotic mass of the spacetime --- that guarantee the existence of trapped and marginally trapped toroids. Explicit solutions are constructed. Section 4 displays graphs of marginally trapped toroids (MTT's); we show, in particular, MTT's that exist both partly in the star and in the surrounding vacuum. We investigate in Section 5 isoperimetric inequalities in order to find those that adequately describe the situation. It seems that the most appropriate are those quantities which are related to the binding energy. The last Section contains conclusions.

\section{Mean curvature of coordinate tori}

We shall assume momentarily static ($K_{ij} = 0$), spherically symmetric, initial data, for a constant density star, with the 3-metric in cylindrical coordinates as
\be
\label{metricss}
ds^2=\Phi^4 (R) \left( dr^2 + r^2d \phi^2 + dz^2 \right),
\ee
where $\Phi (R) = \frac{(1+\beta R^2_0)^{3/2}}{\sqrt{1+\beta R^2}}$ for $R\le R_0$ and $\Phi(R) = 1 + \beta \frac{R^3_0}{R}$ for $R > R_0$. Here $R = \sqrt{r^2 + z^2}$ is the spherical radial coordinate, and $R_0$ is the radius of the star (in flat coordinates).

Assume that we are given a family of coordinate tori, of major radius $a$ and minor radius $b$, with the $z-$axis as the symmetry axis. It is defined by the following formulae:
\begin{enumerate}
\item In region 1 (i.e., $r\le a$): $(r,\phi, z)$, with $r = a - \sqrt{b^2 - z^2}$, the unit normal vector to the torus is given by
\be
(n^i)=\frac{1}{\Phi^2(R)} \left( -\frac{\sqrt{b^2 - z^2}}{b}, 0, \frac{z}{b} \right);
\ee
\item and in region 2 (i.e., $r\ge a$): $(r,\phi, z)$, with $r = a + \sqrt{b^2-z^2}$, we have
\be
(n^i)=\frac{1}{\Phi^2(R)} \left( \frac{\sqrt{b^2 - z^2}}{b}, 0, \frac{z}{b} \right).
\ee
\end{enumerate}

The metric induced on the surface of the torus is given by $ds^2_{ind} = \Phi^4 (R) \left( r^2 d\phi^2 + \frac{b^2}{b^2-z^2} dz^2 \right)$. The mean curvature $H\equiv \nabla_in^i$ is given by following expressions:
\begin{enumerate}
\item in region 1:
\begin{eqnarray}
H&=&\frac{1}{\Phi^2(R)b}\times \nonumber\\ &&
\left[   \frac{a-2\sqrt{b^2-z^2}}{a-\sqrt{b^2-z^2}} +\frac{4\beta \left(   a\sqrt{b^2-z^2}-b^2   \right)        }{1+\beta R^2}  \right] ;
\label{MCss1}
\end{eqnarray}
\item in region 2:
\begin{eqnarray}
H&=&\frac{1}{\Phi^2(R)b}\times \nonumber\\ &&
\left[   \frac{a+2\sqrt{b^2-z^2}}{a+\sqrt{b^2-z^2}} -\frac{4\beta \left(   a\sqrt{b^2-z^2}+b^2   \right)        }{1+\beta R^2}  \right] .
\label{MCss2}
\end{eqnarray}
\end{enumerate}

We assume for convenience that $R_0=1$. Then the conformal factor is given by $\Phi (R) = \frac{(1+\beta )^{3/2}}{\sqrt{1+\beta R^2}}$ for $R \le 1$ and $\Phi (R) = 1 + \frac{\beta}{R}$ for $R > 1$. Note that with this choice, the asymptotic mass of the spacetime becomes $m = 2 \beta$.
 
\section{Analytic constructions: trapped and marginally trapped tori}

We shall present below families of trapped surfaces, including marginally trapped and constant mean curvature tori (CMCT). It is a simple algebraic exercise to show, that if
\be
\beta =\frac{1}{a^2 - b^2},
\label{main1}
\ee
then the mean curvature of the coordinate tori is constant and it is equal to
\begin{eqnarray}
H&=&2\beta \frac{ a^2-2b^2}{  b(1+\beta )^{3}}.
\label{MC1}
\end{eqnarray}
This expression vanishes when $a = \sqrt{2}b$ --- the corresponding torus is marginally trapped. We have $1 \ge a + b = b(1+\sqrt{2})$, since we assume that the tori are inside the star; thus the minimal value of $\beta$ is given by $\beta_{min} =(1 + \sqrt{2})^2 = 3 + 2\sqrt{2}$. If $\beta \ge \beta_{min}$, then there exists a marginally trapped coordinate torus with $b = 1/\sqrt{\beta}$, $a = b\sqrt{2}$ and $a + b\le 1$.

The constant mean curvature tori are trapped if $a = \sqrt{\alpha}b$, where $1 < \sqrt{\alpha} < \sqrt{2}$. In this case Eq.\ (\ref{main1}) yields $\beta = 1/(b^2(\alpha - 1))$; note that $1 \ge a + b = b(1 + \sqrt{\alpha})$. Therefore we have for trapped tori $\beta_{Tmin} = \frac{1 + \sqrt{\alpha}}{-1 + \sqrt{\alpha}}$. This is a decreasing function of $\alpha $; its minimal value is achieved at $\alpha = 2$, that is we get the intuitively obvious estimate: $\beta_{Tmin}>\beta_{min}$. In order to create configurations with constant mean curvature trapped tori one needs a bigger asymptotic mass than for the formation of configurations that contain only marginally trapped toroids. This is proven only for   marginally trapped coordinate tori, but numerical calculations suggest that the only MTT's that exist inside the star are the coordinate tori found in this section. 

In what follows we show how to construct trapped tori, that do not have constant mean curvature. We assume $a$ and $b$ satisfy $a + b = 1$; these tori are tangent to the boundary of a star. This assumption is not necessary, but it is makes the calculation simpler. Moreover, since we exclude self-intersecting tori, we must have $b < 1/2$ and $a > 1/2$.
 
Direct calculation gives in region 1
\begin{eqnarray}
\label{derH1}
\frac{dH}{dz}&=& 
 \frac{K }{ \left( a-\sqrt{(a-1)^2-z^2}\right)^2}
\end{eqnarray}
and in region 2
\begin{eqnarray}
\label{derH2}
\frac{dH}{dz}&=& \frac{-K }{ \left( a+\sqrt{(a-1)^2-z^2}\right)^2}.
\end{eqnarray}
Here $K \equiv \frac{a z \left( -1 + \beta \left( 2a - 1 \right) \right)}{(a - 1)(1 + \beta)^3\sqrt{(a-1)^2-z^2} } $. Note that if $\beta > \frac{1}{2a - 1}$, then --- assuming that $z>0$ --- the mean curvature $H$ is a monotonically decreasing function of $z$ in region 1, and it is a monotonically increasing function in region 2. Note also that $H$ is an even function of $z$. Therefore, in this case it suffices to prescribe nonpositive values of the mean curvature at two points; those farthest and closest to the center. That guarantees that the torus is trapped. 

\section{Numerics: marginally trapped toroidal surfaces}
 
In the following we use standard toroidal coordinates $(\sigma, \tau, \phi)$ \cite{KK}. They are related to the Cartesian coordinates $(x,y,z)$ by $x  = \frac{c \sinh \tau \cos \phi}{\cosh \tau - \cos \sigma}$, $y  =  \frac{c \sinh \tau \sin \phi}{\cosh \tau - \cos \sigma}$ and $z  =  \frac{c \sin \sigma}{\cosh \tau - \cos \sigma}$; here $- \pi \le \sigma \le \pi$, $\tau \ge 0$, $0 \le \phi < 2 \pi$ and $c > 0$ is a radius of the circle in the $z = 0$ plane corresponding to $\tau = \infty$. The metric can be expressed as $ ds^2 = \Phi^4 \frac{c^2}{(\cosh \tau - \cos \sigma)^2} \left( d \sigma^2 + d \tau^2 + \sinh^2 \tau d \phi^2 \right).$ We parametrize a toroidal surface $S$ by $\tau = f(\sigma)$. The mean curvature of $S$ reads
\begin{eqnarray}
\label{rownanie}
H & = & \left. \frac{(\cosh \tau - \cos \sigma)^3}{c \Phi^6(\sigma,\tau) \sinh \tau} \right|_{\tau = f(\sigma)} \times \\
\nonumber
& & \left\{ \partial_\sigma \left. \left[ \frac{\Phi^4(\sigma,\tau) \sinh \tau f^\prime(\sigma)}{(\cosh \tau - \cos \sigma)^2 \sqrt{1 + f^\prime(\sigma)^2}}  \right] \right|_{\tau = f(\sigma)} - \right. \\
& & \left. \partial_\tau \left. \left[ \frac{\Phi^4(\sigma,\tau) \sinh \tau}{(\cosh \tau - \cos \sigma)^2 \sqrt{1 + f^\prime(\sigma)^2}} \right] \right|_{\tau = f(\sigma)} \right\}.
\nonumber
\end{eqnarray} 
The constant mean curvature tori (CMCT) obtained in the previous section can be easily recovered by setting $\Phi = (1 + \beta)^{3/2}/\sqrt{1+\beta R^2}$ and $c = 1/\sqrt{\beta}$. All tori of constant $H$ are then given by $\tau = \mathrm{const}$; the corresponding mean curvature reads
\begin{equation}
H = \frac{\sqrt{\beta } (\cosh (2 \tau )-3) \text{csch}(\tau )}{(\beta +1)^3}.
\end{equation}
One obtains $H = 0$ by setting $\tau =\tau_0 = \log (1 + \sqrt{2})= \mathrm{arccosh}(3)/2$. Fig.\ 1 depicts four such marginally trapped tori that are contained entirely within the constant density star and that exist for  $\beta \ge \beta_0 \equiv 3+2\sqrt{2}$.

Let us consider a family of CMCT's in the vicinity of $\tau_0$. Their area is given by $A_\tau = \frac{\pi^2 (1+\beta)^6 \sinh\tau}{\beta \cosh^2\tau}$. The first derivative $\frac{dA_\tau}{d\tau}$ vanishes at $\tau = \tau_0$ and the second derivative $\frac{d^2 A_\tau}{d \tau^2}|_{\tau=\tau_0} = -\frac{\pi^2(1+\beta)^6}{\beta}$ is strictly negative at $\tau = \tau_0$. This deformation locally maximizes the torus area. Therefore, when $\tau = \tau_0$, the marginally trapped torus is not stable. This agrees with the result of Galloway, who proved that a minimal surface of toroidal topology can be stable only if the three dimensional scalar curvature vanishes along the toroid, and the toroid is totally geodesic [8].

Several MTT's are depictured in Fig.\ 1 for $\beta \ge 3 + 2\sqrt{2}$. They can be obtained numerically also for $\beta < 3 + 2\sqrt{2}$. Examples of such MTT's are shown in Fig.\ 2. In this case only a part of the surface is contained within the star. We compute MTT's by solving the equation $H = 0$, where $H$ is given by Eq.\ (\ref{rownanie}), for the function $f(\sigma)$. Assuming equatorial symmetry we restrict ourselves to the interval $[0,\pi]$ with the boundary conditions $f^\prime(0) = f^\prime(\pi) = 0$. In practice, solutions depicted in Fig.\ 2 are obtained by gluing solutions corresponding to $\Phi = (1 + \beta)^{3/2}/\sqrt{1+\beta R^2}$ inside the star and those corresponding to the vacuum region ($R>1$) where we put $\Phi = 1 + \beta/R$. There is an interesting behaviour of these MTT's. They cross the sphere $R=1$ and go outward until a value $\beta_m \approx 2.58$, and then they go inward and center along a piece of the sphere $R = 1$. They exist for values of $\beta$ that are arbitrarily close to $\beta =1$ --- when trapped spheres do exist ---  but they are obviously absent at $\beta = 1$, because in such a case the boundary of the star coincides with the minimal surface and no spherical trapped surfaces exist.
  
\begin{figure}[t!]
\includegraphics[width=\columnwidth]{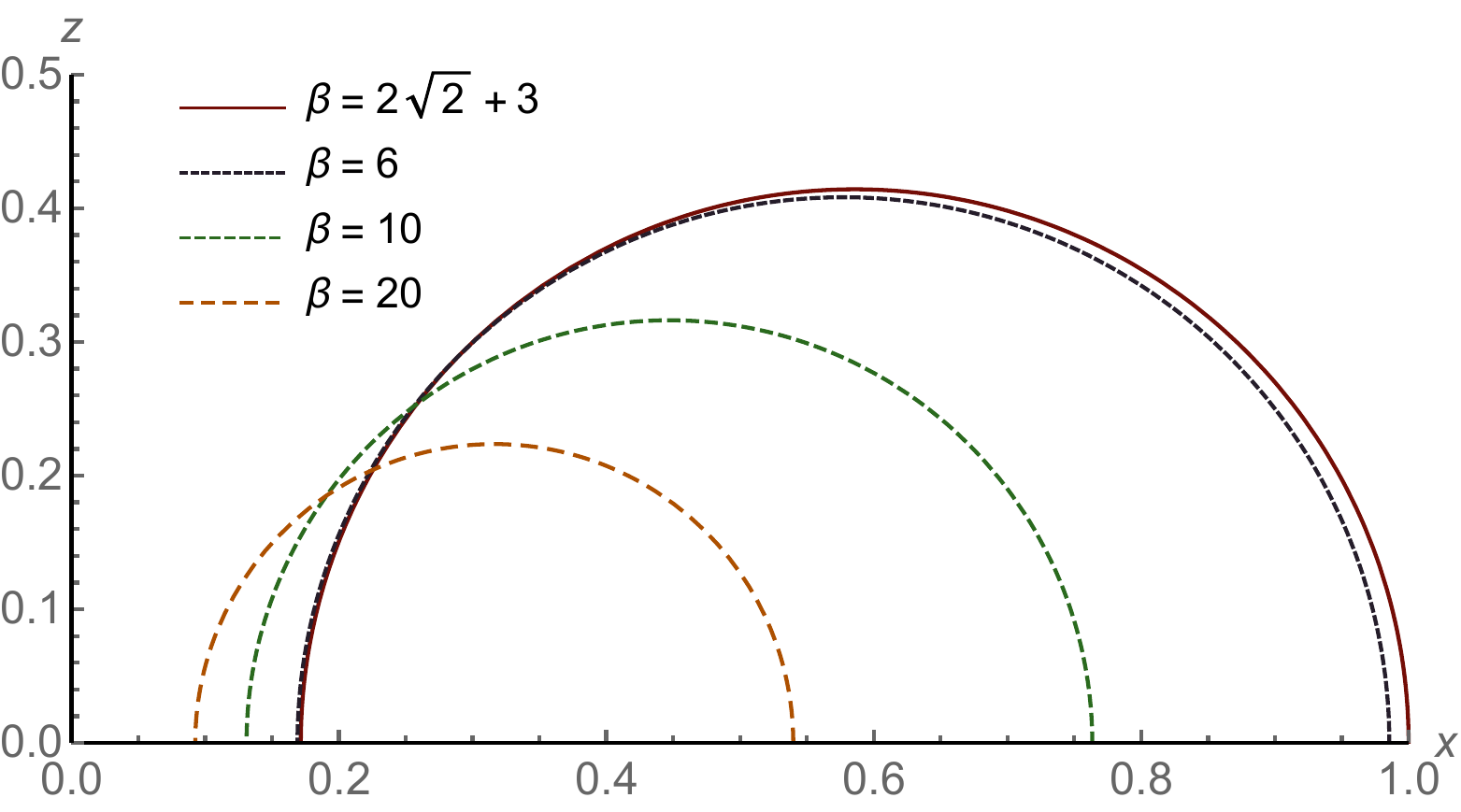}
\caption{\label{rys1}The shapes of MTT's obtained for $\beta =3+2\sqrt{2},    6,  10, 20$. All these surfaces are contained within the star.}
\end{figure}

 \begin{figure}[t!]
 \includegraphics[width=\columnwidth]{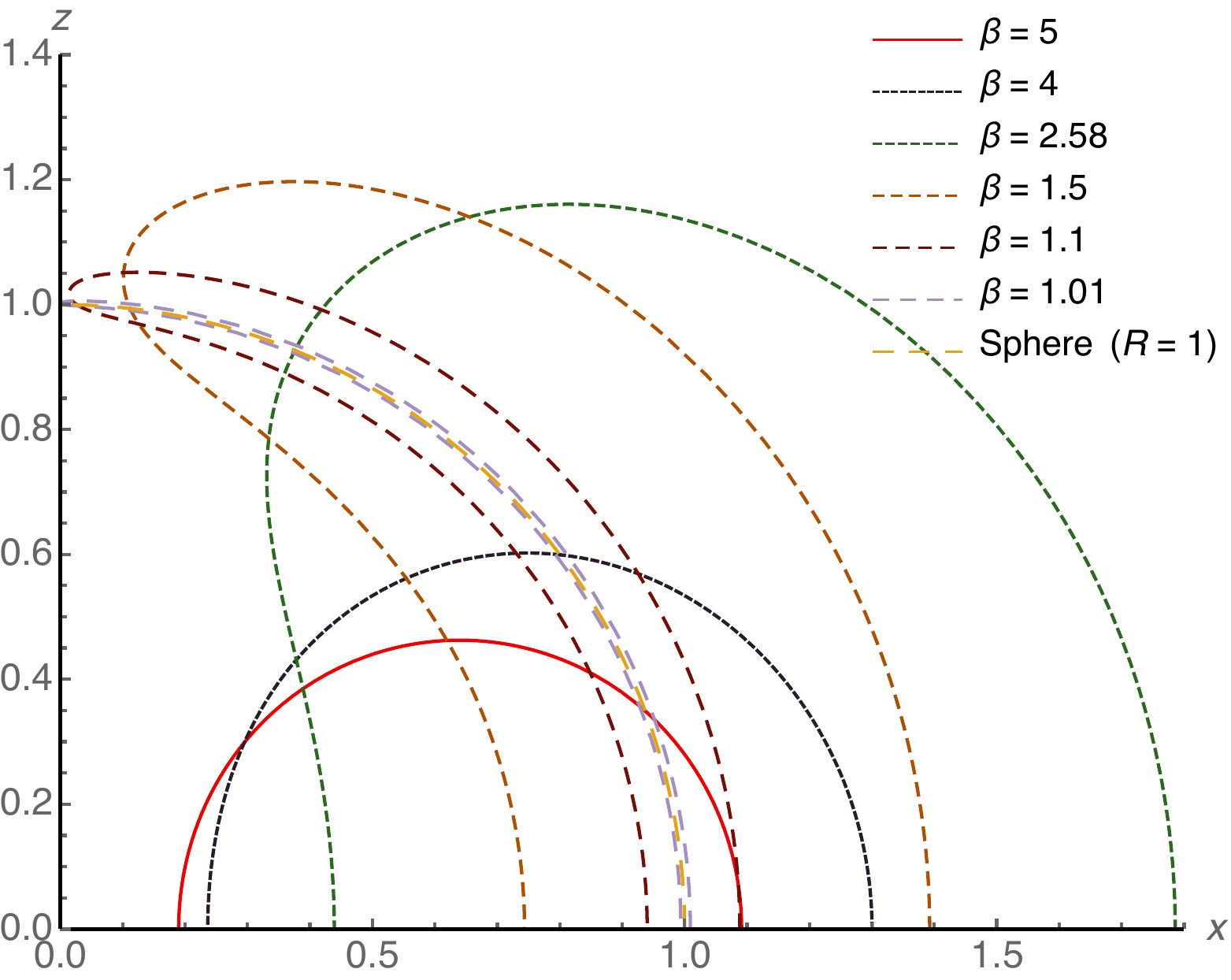}
\caption{\label{rys2}The shapes of MTT's obtained for $\beta = 5, 4, 2.58, 1.5, 1.1, 1.01$.}
\end{figure} 
   
\section{Trapped toroids and isoperimetric inequalities}

We will discuss in this section several inequalities describing toroidal marginally trapped surfaces. Momentarily static initial data, which are given by (\ref{metricss}), correspond to constant mass density
\be
\label{rho}
\rho =\frac{3\beta }{2\pi }\frac{1}{(1+\beta )^6}.
\ee
The corresponding asymptotic mass is $m = 2\beta$ and the rest mass is given by
\begin{eqnarray}
\label{rest}
M & = & \int_VdV\rho \\
& = & \frac{3}{4\sqrt{\beta }}(1+\beta)^3\left( \arctan (\sqrt{\beta}) +\frac{\sqrt{\beta} (\beta - 1)}{(1 + \beta)^2} \right).
\nonumber
\end{eqnarray}
The resulting geometry has been studied in \cite{BMOM}; it turns out that the first spherical minimal surface appears when the parameter $\beta =1$, in which case the proper radius $L = \int_0^1dR\Phi^2$ becomes equal to $4M/3$.

The ratio $m/M$ can be bounded for spherical initial data of compact support \cite{BMOM90}, $\frac{m}{M}\le 1-\frac{M}{2L}$, where $L$ is the proper radius of the star. This inequality, that is valid for spherically symmetric, momentarily static initial geometries, is just one of many examples of isoperimetric inequalities. Its right hand side can be treated as a measure of the strength of gravity. It is equal to 0.625, when $\beta = 1$, that is when the first minimal surface is born, and it is equal to $0.545$ at the value $\beta = 3 + 2\sqrt{2}$. It is more instructive, however, to do the straightforward calculation using the formulae for $m$ and $M$. One obtains
\be
\label{ratiomM}
\frac{m}{M} = \frac{8 \beta}{3 (1 + \beta )^3}\left( \frac{\arctan (\sqrt{\beta })}{\sqrt{\beta }} +\frac{(\beta -1)}{(1+\beta )^2} \right)^{-1}.
\ee
One can check that $m/M\rightarrow 1$ when $\beta \rightarrow 0$. At $\beta = 1$ (allowing the appearance of the first spherical apparent horizon) we have $m/M\approx 0.42$ and at $\beta = 3 + 2\sqrt{2}$ (allowing the appearance of the first MTT) we get $m/M\approx 0.083$. The ratio of the two masses monotonically decreases, and for large $\beta$ we obtain small values of $m/M$.

The quantity $\epsilon_m \equiv 1 - \frac{m}{M}$ can be interpreted as the binding energy per unit rest mass. We hypothesise that the following is true:\\
\noindent \textit{Conjecture. If the value of the binding energy per unit mass  $\epsilon_m$ is large enough, that is $m/M$ is small, then trapped surfaces of topologies different than $S^2$ can form.}

The more precise statement would be that: MTT's will exist if $\epsilon_m > 0.5$. The lower bound $\epsilon_m = 0.5$ is saturated by massive spherical shells, when the minimal surface coincides with the shell itself \cite{BMOM90}, and there should be no  MTT's.

In the remainder of this section we shall discuss inequalities of \cite{KHX}, using our explicit solutions. The total rest mass  of a marginally trapped torus  is given by
\begin{eqnarray}
M_T & = & \int_V \rho dV= \rho \times \mbox{Volume} \\
& = & 3\beta (1+\beta)^3 \int_{-b}^b dz  \int_{a-\sqrt{b^2-z^2}}^{a+\sqrt{b^2-z^2}} \frac{r dr}{(1+\beta(r^2+z^2))^{3}}\nonumber \\
& = &\frac{3(1+\beta )^3}{4\sqrt{\beta} }\int_{-1}^1 dx (I^{-2}_- -I^{-2}_+).
\nonumber
\end{eqnarray}
Here $x = z \sqrt{\beta}$ and $I_{\pm}=1+\beta(a^2 + b^2 \pm 2 a \sqrt{b^2 - z^2})= 4 \pm 2 \sqrt{2} \sqrt{1 - x^2}$.

The circumferential radius $L_r = r \Phi^2$ through a marginally trapped torus is maximized at $r = 1/\sqrt{\beta}$, $z = 0$ and it has a maximal length equal to $L_r =\frac{ (1+\beta )^3}{2\sqrt{\beta }}$.

The Khuri-Xie \cite[Eqn.(5.3), Theorem 5.1 i)]{KHX} bound expressed in terms of the maximal length of the circumferential radius $L_r$ reads now $\frac{2M_T}{\pi L_r} \le 1$. We easily get that the left hand side is equal to $\frac{3}{\pi} \int_{-1}^1 dx (I^{-2}_- - I^{-2}_+) = \frac{3}{4}$. It is remarkable that all marginally trapped tori satisfy this bound with the same coefficient $\frac{3}{4}$. 
 
On the other hand we did not find trapped tori that satisfy the sufficient condition \cite[Eqn.(5.5), Theorem 5.1 iii)]{KHX}. 
 
\section{Summary}

We have constructed explicit and numerical families of toroidal trapped surfaces ($H<0$) and marginally trapped toroids ($H=0$). They constitute the first explicit examples of unstable MTT's. The analytic MTT's show a remarkable universality --- the ratio $M_T/L_r$ is the same for all of them.

We constructed solutions for the fixed orientation of the coordinate system, but our initial data are spherically symmetric, which means that rotated coordinate systems also possess their own MTT's. Thus there exists infinite number of MTT's in our set of initial data. MTT's exist in the volume contained beneath the outermost minimal sphere, which is located at $R=\beta$. We conjecture that they can appear for systems with sufficiently large binding energies.

\section*{Acknowledgments}

Part of this work was done while E.\ Malec and N.\ \'{O} Murchadha were visiting the Shanghai Center for Mathematical Sciences and the School of Mathematical Sciences, Fudan University. They would like to thank these institutions for their hospitality and financial support. P.\ Mach acknowledges the support of the NCN grant DEC-2012/06/A/ST2/00397. N.\ Xie is partially supported by the National Science Foundation of China (Grants 11671089, 11421061).

\end{document}